# A Piezoelectric, Strain-Controlled Antiferromagnetic Memory Insensitive to Magnetic Fields

Han Yan[1,#], Zexin Feng[1,#], Shunli Shang[2], Xiaoning Wang[1], Zexiang Hu[1], Jinhua Wang[3], Zengwei Zhu[3], Hui Wang[1], Zuhuang Chen[4], Hui Hua[1], Wenkuo Lu[1], Jingmin Wang[1], Peixin Qin[1], Huixin Guo[1], Xiaorong Zhou[1], Zhaoguogang Leng[1], Zikui Liu[2], Chengbao Jiang[1], Michael Coey[1,5], Zhiqi Liu[1,*]

1. School of Materials Science and Engineering, Beihang University, Beijing 100191, China
2. Department of Materials Science and Engineering, The Pennsylvania State University, University Park, Pennsylvania 16802, USA
3. Wuhan National High Magnetic Field Center and School of Physics, Huazhong University of Science and Technology, Wuhan 430074, China
4. School of Materials Science and Engineering, Harbin Institute of Technology, Shenzhen 518055, China
5. Department of Pure and Applied Physics, Trinity College, Dublin 2, Ireland

[#]These authors contributed equally to this work.

*email: zhiqi@buaa.edu.cn

**Spintronic devices based on antiferromagnetic (AFM) materials hold the promise of fast switching speeds and robustness against magnetic fields[1-3]. Different device concepts have been predicted[4,5] and experimentally demonstrated, such as low-temperature AFM tunnel junctions that operate as spin-valves[6], or room-temperature AFM memory, for which either thermal heating in combination with magnetic fields[7], or Néel spin-orbit torque[8] is used for the information writing process. On the other hand, piezoelectric materials were employed to control magnetism by electric fields in multiferroic heterostructures[9-12], which suppresses Joule heating caused by switching currents and may enable low energy-consuming electronic devices. Here, we combine the two material classes to explore changes of the resistance of the high-Néel-temperature antiferromagnet MnPt induced by piezoelectric strain. We find two non-volatile resistance states at room temperature and zero electric field, which are stable in magnetic fields up to 60 T. Furthermore, the strain-induced resistance switching process is insensitive to magnetic fields. Integration in a tunnel junction can further amplify the electroresistance. The tunneling anisotropic magnetoresistance reaches ~11.2% at room temperature. Overall, we demonstrate a piezoelectric, strain-controlled AFM memory which is fully operational in strong magnetic fields and has potential for low-energy and high-density memory applications.**

MnPt is a collinear AFM intermetallic with a Néel temperature of ~975 K[13], which has been extensively employed as an exchange bias material[14]. The high Néel temperature suggests that magnetic coupling between two antiparallel sublattices is very strong[15] and should be exceptionally insensitive to external magnetic fields. To realize possible piezoelectric control of this compound, we have fabricated MnPt thin films on piezoelectric $0.72PbMg_{1/3}Nb_{2/3}O_3$–$0.28PbTiO_3$ (PMN-PT) single-crystal substrates. Figure 1a shows a transmission electron microscopy image of a 37-nm-thick MnPt/PMN-PT heterostructure, which reveals a reasonably sharp interface between the film and the PMN-PT substrate. X-ray diffraction measurements indicate that the films are textured with preferred (101) and (001) orientations (Fig. 1b) (Supplementary Note 1). The absence of the (001) Bragg peak is possibly due to the low growth temperature of MnPt and the resulting imperfect structural order. The composition of the films was determined to be $Mn_{48.3}Pt_{51.7}$ (±2.0 at%) by energy-dispersive X-ray spectroscopy.

Preliminary magnetic measurements performed in a vibrating sample magnetometer indicate the absence of any magnetic hysteresis loop in MnPt films. We could not detect any magnetic signal (Fig. 1c & d) within the detection limit of ~0.005 $\mu_B$/atom[16]. These results are consistent with the expected AFM nature of MnPt. To confirm the AFM order of the MnPt film, we prepared a second MnPt thin film and then deposited a 2.5 nm soft ferromagnetic $Co_{90}Fe_{10}$ film with an intrinsic coercivity $\mu_0 H_c$ of 1.0-1.5 mT and finally a 2 nm Pt capping layer to form a Pt(2)/$Co_{90}Fe_{10}$(2.5)/MnPt(37)/PMN-PT (thickness in nm) heterostructure. A large exchange bias $\mu_0 H_{EB}$ of ~24.5 mT was found and the coercivity $\mu_0 H_c$ of $Co_{90}Fe_{10}$ was greatly increased to ~8.0 mT (Fig. 1e) at room temperature. After cycling the magnetic field between 100 and -100 mT for 500 full cycles at a rate of 5 mT/s for this sample, the exchange bias remains stable. Such a robust exchange bias corroborates the AFM order of the MnPt film.

Magnetotransport measurements were carried out for the MnPt/PMN-PT heterostructure at room temperature in order to explore the magnetic field sensitivity of the resistance of the MnPt film. The measurement geometry is schematized in Fig. 2a. As shown in Fig. 2b, the room-temperature four-probe resistance of the film is ~15.4 Ω, corresponding to a resistivity of ~183 μΩ·cm, which is comparable to the room-temperature resistivity of bulk MnPt[17]. More importantly, both the out-of-plane magnetoresistance and the in-plane, parallel magnetoresistance are very small up to field strength of 9 T, on the order of ~0.01% (Fig. 2c).

Next, we examined the effect of an electric field $E_G$ applied via the bottom gate across the PMN-PT substrate on the electrical properties of the MnPt thin film. The measurement geometry is illustrated in Fig. 2d. The resistance of MnPt is very sensitive to $E_G$, and the $E_G$-dependent resistance exhibits a hysteretic and asymmetric butterfly shape (Fig. 2e). $E_G$ modulates the resistance by up to ~2%. More importantly, at $E_G = 0$ kV/cm, there are two distinct resistance states – a high-resistance state and a low-resistance state in the $R(E_G)$ curve (Fig. 2e).

The electroresistance modulation is unlikely to be due to electrostatic carrier injection, because a negative $E_G$ should correspond to an increase of the MnPt resistance in the electrostatic modulation case with our measurement geometry (Supplementary Figure 1 and Supplementary Note 2). Moreover, the Thomas-Fermi electrostatic screening length for an intermetallic alloy with a high carrier density is very short, typically on the order of angstrom[18], and is much smaller than our film thickness. Instead, the $R(E_G)$ curve (Fig. 2e) resembles the $E_G$-dependent strain curve in (001)-oriented PMN-PT substrates[19,20], which was found to be due to 109° ferroelastic domain switching[19-21]. The peaks of the gate current in Fig. 2f are evidences of polarization switching in PMN-PT.

Following this line of thought, we measured the $E_G$-induced lattice change of PMN-PT by x-ray diffraction. We studied sequentially the (004) diffraction peak of PMN-PT at $E_G = 6.7$, 0 and -6.7 kV/cm. The peak shift can be clearly seen in Fig. 3a. Compared with $E_G = 0$ kV/cm, the out-of-plane lattice constant of PMN-PT is elongated at both $E_G = \pm 6.7$ kV/cm, which corresponds to an in-plane compressive strain induced by $E_G$. The out-of-plane strain $\varepsilon_c$ is determined to be ~0.08% and ~0.15% for $E_G = +6.7$ and -6.7 kV/cm, respectively. By using the measured value of the Poisson's ratio $\nu \approx 0.43$ for PMN-PT[22], we calculated the in-plane strain $\varepsilon_a$ to be ~-0.05% at $E_G = +6.7$ kV/cm and ~-0.1% at $E_G = -6.7$ kV/cm. This is consistent with the asymmetric feature of the $E_G$-dependent strain in (001)-oriented PMN-PT single crystals, which has been discussed previoulsy[19,20].

To better understand how the strain changes the resistivity of MnPt in the MnPt/PMN-PT heterostructure as shown in Fig. 2e, density functional theory calculations were performed. In an unstrained primitive cell with two Mn and two Pt atoms (Supplementary Figure 2 and Supplementary Note 3), the two Mn magnetic moments (~3.67 $\mu_B$) are antiparallel to each other.

Due to hybridization of the *d* electrons of Mn and Pt atoms, Pt carries a small magnetic moment of ~0.11 $\mu_B$ and the spins of the two Pt atoms are opposite as well. It was found that both the band structure and the density of states are almost unchanged on applying an in-plane compressive strain of 0.1% (see Supplementary Figure 2). In addition, Hall measurements reveal that the carrier density remains the same in the different resistance states (see Supplementary Figure 3).

To detect whether strain modifies the magnetic properties of AFM MnPt in the MnPt/PMN-PT heterostructure, we examined the effect of $E_G$ on the exchange bias (Fig. 1e) that MnPt exerts on the $Co_{90}Fe_{10}$ overlayer in the Pt(2)/$Co_{90}Fe_{10}$(2.5)/MnPt(37)/PMN-PT heterostructure. Similar to previous reports on the exchange bias that can be electrically controlled using either a magnetoelectric or a piezoelectric material[23-25], it was found that both $\mu_0 H_{EB}$ and $\mu_0 H_c$ are very sensitive to $E_G$ in our system and they can be reversibly switched by positive/negative electric-field pulses applied across the piezoelectric substrate (Fig.3b). In the non-volatile, low-resistance state of MnPt, $\mu_0 H_{EB}$ and $\mu_0 H_c$ are increased to 30.5±0.5 and 9.4±0.2 mT, respectively. As the piezoelectric strain does not alter the magnetic properties of the ferromagnetic $Co_{90}Fe_{10}$ overlayer (see Supplementary Figure 4), such an enhancement of the exchange bias indicates that magnetic properties of MnPt are significantly changed.

The largest exchange coupling in the $Co_{90}Fe_{10}$/MnPt system achieved so far, has been in highly-ordered (001)-oriented MnPt epitaxial films[26]. In addition, neutron diffraction studies indicate that the AFM axis is along the *c* in MnPt below 750 K[27]. Therefore, there should be at least two orientations for the AFM spin axis in our textured MnPt films: one is out-of-plane for (001)-oriented grains and the other is at 45° relative to the normal to the sample surface for (101)-oriented grains. The mechanism for the large exchange coupling[26] between $Co_{90}Fe_{10}$ and (001)-oriented MnPt is still not clear. Yet, the increase could be due to the rotation of the AFM axis in (101)-oriented grains towards the normal to the sample surface under the biaxial compressive piezoelectric strain. The hypothetical mechanism is illustrated in Fig. 3c & d. As a result, the enhanced perpendicular geometry between the AFM axis and the current yields the low-resistance state of the MnPt film (Fig. 2e) due to the anisotropic magnetoresistance effect in antiferromagnets[3].

To verify this picture, we fabricated highly-ordered (001)-oriented MnPt films on 3-nm-thick MgAl$_2$O$_4$ buffer layers grown on PMN-PT substrates. In this case, the piezoelectric strain induced by $E_G$ affects neither the resistivity of fully (001)-oriented MnPt nor its exchange bias with a Co$_{90}$Fe$_{10}$ overlayer (see Supplementary Figures 5-8 and Supplementary Note 4). Hence, the spin axis of (001)-oriented MnPt is stable under the biaxial compressive piezoelectric strain and these results support an electroresistance modulation in textured MnPt/PMN-PT heterostructures that originates from the rotation of the AFM spin axis in (101)-oriented grains.

Ferroelectric oxides tend to crack on repeated cycling of the electric field due to stress at the domain boundaries between electrically switchable domains and domains pinned by surface defects[28]. We therefore conducted the electroresistance modulation measurements by using a unipolar switching approach in order to avoid cracking[29]. As seen in Fig. 4a, the unipolar modulation can produce a high-resistance state and a low-resistance state similar to those in Fig. 2e. The resistance difference between the two non-volatile resistance states is ~252 mΩ, which is one order of magnitude larger than that of recently reported AFM memories[7,8].

We can also switch the resistance states by 50-ms pulses of $E_G$ = +1.87 and -6.67 kV/cm at zero field, but also at 9 and 14 T (Fig. 4b). We tested the strain-controlled memory device for 10 days in air with a daily temperature variation of ~3 K. The resistance state was stable with a resistance fluctuation of ~0.05% (Fig. 4c), which is much smaller than the non-volatile resistance modulation, ~1.66% (Fig. 4a). Furthermore, we applied an out-of-plane pulsed field up to 60 T to the low-resistance state and observed a resistance variation of ~0.1% (Fig. 4d), which indicates the information in this memory is not destroyed even by a 60 T magnetic field. In contract, the magnetoresistance of AFM Mn$_3$Pt, which has a lower Néel temperature of ~475 K and is of interest due to the novel anomalous Hall effect[30,31], reaches ~1.8% at 60 T (Fig. 4e).

Finally, we fabricated a tunneling anisotropic magnetoresistance junction[6] based on a textured MnPt/PMN-PT heterostructure by depositing a 2-nm-thick MgAl$_2$O$_4$ tunnel barrier and a 10-nm-thick Pt top layer, which was patterned into microelectrodes with a diameter of 100 μm (schematized in Fig. 5a). Figure 5b shows the room-temperature $E_G$ dependence of the two-probe tunneling anisotropic magnetoresistance in the Pt(10)/MAO(2)/MnPt(37)/PMN-PT (thickness in nm) device. The electroresistance ratio is enhanced up to ~11.2% and the non-volatile electroresistance ratio reaches ~8.7%. The effect could be further amplified by sophisticated microstructure patterning and four-probe measurements to exclude contact resistance. The benefit of using piezoelectric strain control of AFM materials in the perpendicular tunnel junction geometry is significant for high-density memory applications.

In summary, we have achieved a non-volatile electroresistance modulation in a high-Néel-temperature intermetallic compound via piezoelectric strain, thus demonstrating an antiferromagnetic memory, in which both the information writing and storage are robust under extremely large magnetic fields. By contrast, both the information writing process and the resistance states of the CuMnAs-based AFM memory operated by Néel spin-orbit torque are largely affected by a field of 12 T[8]. More importantly, the electroresistance modulation can be amplified by tunneling anisotropic magnetoresistance in a room-temperature tunnel junction. Besides the low power consumption, the piezoelectric strain approach could be applied to intriguing physical phenomena in antiferromagnets, such as magnetic Weyl fermions with broken time reversal symmetry [32], to form a new subfield: antiferromagnetic piezospintronics.

# Figure captions

**Figure 1 | Structure & antiferromagnetic order of MnPt. a,** Transmission electron microscopy image of an interfacial region of of a 37-nm-thick MnPt/0.72PbMg$_{1/3}$Nb$_{2/3}$O$_3$–0.28PbTiO$_3$ (PMN-PT) heterostructure. **b,** X-ray diffraction of the MnPt/PMN-PT heterostructure, **c,** X-ray absorption and **d,** X-ray magnetic circular dichroism spectra collected around the Mn $L_{2,3}$ edges at room temperature of the MnPt/PMN-PT heterostructure. **e,** Exchange bias in a Pt(2)/Co$_{90}$Fe$_{10}$(2.5)/MnPt(37)/PMN-PT heterostructure (thickness in nm) that confirms the antiferromagnetic order of the MnPt film.

**Figure 2 | Magneto- & electro-transport properties of the MnPt film at room temperature. a,** Schematic of the magnetotransport measurement geometry for the MnPt/0.72PbMg$_{1/3}$Nb$_{2/3}$O$_3$–0.28PbTiO$_3$ (PMN-PT) heterostructure. **b,** Out-of-plane and in-plane parallel Magnetoresistance. **c,** Magnetoresistance of 9 T depending on the angle between the magnetic field and the measuring current (100 μA). **d,** Illustration of the electric field ($E_G$) gating geometry. **e,** $E_G$-dependent resistance (measured by 100 μA) of the MnPt film. **f,** $E_G$-dependent gate current of the PMN-PT substrate. The curved arrows and the numbers represent the measurement procedure.

**Figure 3 | Possible mechanism for the electroresistance modulation in the MnPt film. a,** Room-temperature X-ray diffraction of the (004) peak of 0.72PbMg$_{1/3}$Nb$_{2/3}$O$_3$–0.28PbTiO$_3$ (PMN-PT) in the MnPt/PMN-PT heterostructure under different electric field ($E_G$), which was changed from +6.7 to 0 and finally to -6.7 kV/cm. **b,** Exchange bias of the Pt(2)/Co$_{90}$Fe$_{10}$(2.5)/MnPt(37)/PMN-PT heterostructure under different resistance states of MnPt achieved by sequential positive/negative electric field pulses, suggesting the significant piezoelectric strain modulation of magnetic properties of antiferromagnetic MnPt. **c, & d,** Schematic of the antiferromagnetic spin axis distribution at the high-resistance state and low-resistance state of the MnPt/PMN-PT heterostructure, respectively.

**Figure 4 | Piezoelectric strain-controlled antiferromagnetic memory based on the MnPt/0.72PbMg$_{1/3}$Nb$_{2/3}$O$_3$–0.28PbTiO$_3$ heterostructure insensitive to magnetic fields. a,** Unipolar resistance switching of the MnPt film at room temperature by the electric field ($E_G$). The curved arrows and the numbers represent the measurement procedure. **b,** The high-resistance state and the low-resistance state realized by pulses of $E_G$ = +1.87 and -6.67 kV/cm at room temperature, respectively, under 0, 9 and 14 T. The area highlighted in yellow on the right part of the figure stands for the existence of a 9 T or a 14 T magnetic field. **c,** Stability of the low-resistance state (LRS) as a function of time while the memory device is exposed in air. The resistance of the LRS was read by a 100 μA current pulse every 60 s for more than 10 days. **d,** Magnetic field dependence of the LRS under a pulsed magnetic field up to 60 T at room temperature. **e,** Room-temperature magnetoresistance of MnPt and Mn$_3$Pt films up to 60 T.

**Figure 5 | Piezoelectric strain-controlled room-temperature antiferromagnetic tunnel junction. a,** Schematic of the junction structure built on 0.72PbMg$_{1/3}$Nb$_{2/3}$O$_3$–0.28PbTiO$_3$ (PMN-PT) and the measurement geometry. **b,** Electric field ($E_G$) dependence of the two-probe tunneling resistance for a Pt(10)/MgAl$_2$O$_4$(2)/MnPt(37)/PMN-PT (thickness in nm) device.